\documentclass[12pt]{article}

\usepackage{amsmath,amssymb,amsfonts}
\usepackage{bm}
\usepackage{geometry}
\usepackage{graphicx}
\usepackage{cite}
\newcommand{\be}{\begin{equation}}
\newcommand{\ee}{\end{equation}}
\newcommand{\bea}{\begin{eqnarray}}
\newcommand{\eea}{\end{eqnarray}}

\begin{document}

\title{Gauge Theory of Gravity and the AdS/CFT Correspondence}

\author{
Takeshi Fukuyama\\[2mm]
{\small Research Center for Nuclear Physics (RCNP), Osaka University, Ibaraki, Osaka 567--0047, Japan}
}

\maketitle
\begin{abstract}
We discuss the AdS/CFT correspondence from the viewpoint of the gauge-theoretic formulation of gravity, in which gravity is interpreted as a broken phase of conformal gauge symmetry.

In the AdS$_2$/CFT$_1$ case, we show that the Schwarzian derivative naturally emerges from the boundary extrinsic curvature of AdS$_2$ geometry. The relation between the bulk Liouville geometry and the boundary projective structure is clarified. We further discuss the distinction between the bulk conformal gauge algebra with vanishing central extension and the emergent boundary Virasoro structure with nonvanishing central charge.

We then investigate the possible structure of the AdS$_4$/CFT$_3$ correspondence, which is directly related to the original four-dimensional formulation of gravity as a broken phase of conformal gauge symmetry. In this framework, the Einstein--Hilbert action with cosmological constant emerges together with a total derivative term. We argue that this structure induces the boundary gravitational Chern--Simons term, whose variation leads naturally to the Cotton tensor. The Cotton tensor is interpreted as the fundamental conformal invariant associated with the residual boundary conformal geometry, playing a role analogous to that of the Schwarzian derivative in AdS$_2$/CFT$_1$.

We also discuss the qualitative difference between AdS$_4$/CFT$_3$ and AdS$_5$/CFT$_4$. While the former appears naturally connected with gravity arising from conformal symmetry breaking, the latter may require genuinely higher-dimensional, string-inspired structures beyond the four-dimensional conformal gauge framework. These observations suggest a unified geometrical interpretation of holography in terms of boundary remnants of broken conformal gauge symmetry.

\end{abstract}

\section{Introduction}

The AdS/CFT correspondence has played a central role in modern studies of quantum
gravity and gauge theories \cite{Maldacena,Witten}. In its standard form, anti-deSitter gravity on $d+1$-dimensions AdS$_{d+1}$ is related to a conformal field theory CFT$_d$ on the
$d$-dimensional boundary.

Despite its remarkable success, the geometrical origin of the boundary conformal
symmetry is still not fully understood.
In particular, it is important to clarify why conformal structures naturally emerge
at the boundary of AdS gravity and how they are related to the bulk gravitational
geometry itself.

In the present work, we discuss this problem from the viewpoint of the
gauge-theoretic formulation of gravity developed previously by us \cite{FK1,FK2}, in which gravity is interpreted as a broken phase of
conformal gauge symmetry.
In this formulation, the fundamental symmetry is $SO(2,2)$ 
in two dimensions and $SO(4,2)$ in four dimensions, which are spontaneously broken to
$SO(1,2)$ and $SO(3,2)$, respectively.

The essential viewpoint of the present paper is that the boundary conformal
structure appearing in the AdS/CFT correspondence should be interpreted as a
residual manifestation of the original broken conformal gauge symmetry.

In the AdS$_2$/CFT$_1$ case \cite{AP,MSY}, we show explicitly that the
Schwarzian derivative naturally emerges from the boundary extrinsic curvature
of AdS$_2$ geometry.
The relation between the bulk Liouville geometry and the boundary projective
structure is clarified.
We further discuss the relation between the original bulk conformal gauge algebra,
which has no central extension, and the emergent boundary Virasoro structure
with nonvanishing central charge \cite{BH}.

The AdS$_4$/CFT$_3$ correspondence \cite{ABJM} is of particular
interest from this viewpoint, since it is directly related to the
original four-dimensional formulation of gravity as a broken phase
of conformal gauge symmetry.
A central observation is that, in the four-dimensional gauge-theoretic formulation
of gravity, the Einstein-Hilbert action with cosmological constant emerges together
with the total derivative term $\partial_\mu K^\mu$.
We argue that this total derivative structure naturally induces the boundary
gravitational Chern--Simons term, whose variation leads to the Cotton tensor
\cite{DJT, Garcia}.

The Cotton tensor plays the role of the fundamental conformal invariant on the
three-dimensional boundary, in close analogy with the role of the Schwarzian
derivative in the AdS$_2$/CFT$_1$ correspondence.
From this viewpoint, the correspondences 
\be
K-\frac1l
\longleftrightarrow
\{f,u\}
\label{Sch}
\ee
in two dimensions and 
\be
\nabla_k K_{ij}-\nabla_jK_{ik} \longleftrightarrow C_{ijk}
\label{Cotton}
\ee
in four dimensions may be regarded as parallel manifestations of residual boundary structures
originating from broken conformal gauge symmetry.

This viewpoint further suggests that different asymptotically AdS geometries may
correspond to different boundary conformal structures characterized by the
Schwarzian derivative in AdS$_2$/CFT$_1$ and by the Cotton tensor in
AdS$_4$/CFT$_3$.
In particular, asymptotically AdS$_4$ geometries with nontrivial boundary Cotton
tensor may correspond to nontrivial conformal excitations of the boundary theory.
We further discuss the qualitative difference between AdS$_4$/CFT$_3$ and
AdS$_5$/CFT$_4$ within the gauge-theoretic formulation of gravity.
In the straightforward five-dimensional extension of the gauge-theoretic
construction, higher-curvature structures arise naturally, and the ordinary
Einstein--Hilbert action does not emerge in the same manner as in four dimensions.
This suggests that the gauge-theoretic origin underlying AdS$_5$/CFT$_4$
may be qualitatively different from that of AdS$_4$/CFT$_3$.

This paper is organized as follows.
In Sec.~2, we discuss the AdS$_2$/CFT$_1$ correspondence and the relation
between boundary conditions and central extension.
Sec.~3 is devoted to the discussion of the AdS$_4$/CFT$_3$ correspondence.
We then discuss the AdS$_5$/CFT$_4$ correspondence in Sec.~4.
For completeness, a brief summary of the de Sitter invariant formulation
of gravity is given in Appendix~A.

\section{AdS$_2$/CFT$_1$ Correspondence from Gauge Theory of Gravity}
The Lagrangian is given by the well-known Jackiw--Teitelboim action,
Eq.~(\ref{JT2}) \cite{JT,Teitelboim},
whose gauge-theoretic formulation is summarized in Appendix A.
\subsection{AdS$_2$ Geometry}
We consider the AdS$_2$ metric
\begin{equation}
ds^2=\frac{l^2}{z^2}(d\tau^2+dz^2).
\end{equation}
The boundary curve is parametrized by
\begin{equation}
\tau=f(u),\qquad z=z(u).
\end{equation}
The induced metric on the boundary is fixed by
\begin{equation}
ds_\partial=\frac{l}{\epsilon}du.
\end{equation}
This condition gives
\begin{equation}
\frac{l^2}{z^2}(f'^2+z'^2)=\frac{l^2}{\epsilon^2}.
\end{equation}
Expanding in powers of $\epsilon$, we obtain
\begin{equation}
z(u)=\epsilon f'(u)+O(\epsilon^3).
\end{equation}

\subsection{Boundary Extrinsic Curvature}

The extrinsic curvature is defined by
\begin{equation}
K=\nabla_\mu n^\mu ,
\end{equation}
where \(n^\mu\) is the outward-pointing unit normal vector
to the boundary trajectory.
Using the unit tangent and normal vectors, we obtain
\begin{equation}
K
=\frac{1}{l}
\frac{
f'(f'^2+z'^2+zz'')-zz'f''
}{
(f'^2+z'^2)^{3/2}
}.
\end{equation}

Substituting
\begin{equation}
z=\epsilon f'+O(\epsilon^3),
\end{equation}
we obtain
\begin{equation}
K
=\frac1l
+\frac{\epsilon^2}{l}
\left[
\frac{f'''}{f'}
-\frac32
\left(\frac{f''}{f'}\right)^2
\right]
+O(\epsilon^4).
\end{equation}

The quantity in the bracket is precisely the Schwarzian derivative
\begin{equation}
\{f,u\}
=\frac{f'''}{f'}
-\frac32
\left(\frac{f''}{f'}\right)^2.
\label{Schwarzian}
\end{equation}

Thus, the Schwarzian action naturally emerges from the boundary geometry of AdS$_2$.

\subsection{Liouville Geometry and Boundary Projective Structure}

In the gauge-theoretic formulation of gravity, the AdS$_2$ geometry may be described by
\begin{equation}
e_{\mu a}=e^\chi\delta_{\mu a}.
\end{equation}

The constant curvature condition leads to the two-dimensional Liouville equation
\begin{equation}
\partial^2\chi\sim e^{2\chi}.
\end{equation}

This Liouville structure describes the bulk conformal geometry.

On the other hand, the Schwarzian derivative is associated with the boundary reparametrization mode and projective geometry.

Thus, the AdS$_2$/CFT$_1$ correspondence may be interpreted as a reduction
\[
\text{bulk conformal geometry}
\rightarrow
\text{boundary projective geometry}.
\]
\subsection{Boundary Symmetry and Central Extension}

An important issue is the relation between the bulk gauge algebra in the
Fukuyama--Kamimura formulation and the boundary conformal algebra.

In the original Fukuyama--Kamimura formulation, the AdS$_2$ gauge algebra
closes without a central extension.  In the notation of Ref.~\cite{FK2},
this may be expressed as
\begin{equation}
[L_m,L_n]
=
(m-n)L_{m+n},
\end{equation}
with no central term.  Thus the central charge of the bulk gauge algebra is
\begin{equation}
c_{\rm bulk}=0 .
\end{equation}

This statement refers to the bulk gauge algebra before imposing asymptotic
AdS boundary conditions.  In the presence of an asymptotic AdS boundary,
the generator must be improved by a surface contribution,
\begin{equation}
\widetilde L_n
=
L_n+Q_n,
\end{equation}
so that its variation is well-defined under the boundary conditions.

The algebra of the improved generators then takes the form
\begin{equation}
[\widetilde L_m,\widetilde L_n]
=
(m-n)\widetilde L_{m+n}
+
K_{m,n},
\end{equation}
where \(K_{m,n}\) is a boundary contribution.

For asymptotic conformal transformations, this boundary contribution becomes
the Virasoro central extension Eq.~(\ref{algebra}).

In the Brown--Henneaux normalization \cite{BH} one obtains Eq.~(\ref{center}).

Thus, the nonvanishing central charge is not a central extension of the
original bulk Fukuyama--Kamimura gauge algebra.  It appears only after
imposing asymptotic AdS boundary conditions and adding the corresponding
surface charges.  In this sense, the boundary Virasoro central charge is an
emergent boundary structure associated with the residual conformal symmetry.

\section{Toward AdS$_4$/CFT$_3$}

The AdS$_4$/CFT$_3$ correspondence is of particular
interest from the viewpoint of conformal symmetry
breaking gravity.

In the four-dimensional case, gravity emerges through the symmetry breaking
\begin{equation}
SO(4,2)\rightarrow SO(3,2).
\end{equation}

The resulting theory becomes Einstein gravity with cosmological constant. See Appendix A.1 and Ref.~\cite{FK1} for more details.

This situation is qualitatively different from the five-dimensional case discussed later.

The boundary of AdS$_4$ is three-dimensional, and the conformal structure is expected to be described by three-dimensional conformal geometry.
To clarify the origin of the boundary conformal structure in the AdS$_4$/CFT$_3$
correspondence, let us reconsider the total derivative term appearing already in
the gauge-theoretic formulation of gravity developed in Ref.~\cite{FK1}.

In four dimensions, the Euler density may be written as (see Eq.~(\ref{K})),
\begin{equation}
E_4
=
\epsilon_{abcd}
\mathring{R}^{ab}\wedge \mathring{R}^{cd},
\end{equation}
which is locally expressed as a total derivative
\begin{equation}
E_4=dQ_3(\omega).
\end{equation}

Here
\begin{equation}
Q_3(\omega)
=
\epsilon_{abcd}
\left(
\omega^{ab}\wedge d\omega^{cd}
+
\frac23
\omega^{ab}\wedge \omega^c{}_e\wedge\omega^{ed}
\right)
\end{equation}
is the gravitational Chern--Simons 3-form.

Thus, the total derivative structure
\begin{equation}
\partial_\mu K^\mu
\end{equation}
appearing in the original gauge-gravity action Eq. (\ref{K}) in Appendix may be interpreted as the
boundary contribution
\begin{equation}
\int_M d^4x\,\partial_\mu K^\mu
=
\int_{\partial M} Q_3 .
\end{equation}

Let $\gamma_{ij}$ denote the induced metric on the three-dimensional boundary.
The corresponding gravitational Chern--Simons action is
\begin{equation}
I_{\rm gCS}[\gamma]
=
\int_{\partial M}
d^3x\,\epsilon^{ijk}
\left(
\Gamma^l{}_{im}\partial_j\Gamma^m{}_{kl}
+
\frac23
\Gamma^l{}_{im}\Gamma^m{}_{jn}\Gamma^n{}_{kl}
\right).
\end{equation}

Its variation gives
\begin{equation}
\delta I_{\rm gCS}
=
\int_{\partial M}
d^3x\sqrt{\gamma}\,
C^{ij}\delta\gamma_{ij},
\end{equation}
where
\begin{equation}
C^{ij}
=
\epsilon^{ikl}
\nabla_k
\left(
R^j{}_l
-\frac14\delta^j_lR
\right)
\end{equation}
is the Cotton tensor \cite{DJT, Garcia}.

Therefore,
\begin{equation}
\frac{\delta}{\delta\gamma_{ij}}
\int_{\partial M}Q_3
\propto
C^{ij}.
\end{equation}

This result suggests that the Cotton tensor naturally emerges from the
boundary variation of the total derivative term already contained in the
gauge-theoretic formulation of gravity.

The structure is analogous to the AdS$_2$/CFT$_1$ correspondence, where the
boundary Schwarzian derivative arises from the expansion of the extrinsic
curvature,
\begin{equation}
K
=
\frac1l
+
\epsilon^2\{f,u\}
+\cdots .
\end{equation}

Thus, the correspondence Eq.~(\ref{Sch})

is generalized in the AdS$_4$/CFT$_3$ case to Eq.~(\ref{Cotton}).

This observation supports the interpretation that the boundary conformal
structure in AdS$_4$/CFT$_3$ arises as a residual manifestation of the
original broken conformal gauge symmetry.
An important property of the Cotton tensor is its conformal invariance.
In three dimensions, the Weyl tensor vanishes identically,
\begin{equation}
W_{ijkl}\equiv0,
\end{equation}
and therefore the Cotton tensor plays the role of the fundamental conformal
curvature.

Under a conformal transformation of the boundary metric,
\begin{equation}
\gamma_{ij}\rightarrow e^{2\sigma(x)}\gamma_{ij},
\end{equation}
the Cotton tensor transforms covariantly and vanishes if and only if the
geometry is conformally flat.

This situation is closely analogous to the role of the Schwarzian derivative
in the AdS$_2$/CFT$_1$ correspondence.

Indeed, the Schwarzian derivative Eq.~(\ref{Schwarzian})
is invariant under the projective transformation
\begin{equation}
f(u)\rightarrow
\frac{af(u)+b}{cf(u)+d},
\qquad
ad-bc\neq0.
\end{equation}

Thus, in the AdS$_2$/CFT$_1$ case, the Schwarzian derivative represents the
fundamental invariant associated with the residual projective structure on the
boundary.

On the other hand, in the AdS$_4$/CFT$_3$ case, the Cotton tensor represents
the fundamental invariant associated with the residual conformal structure on
the three-dimensional boundary.

This correspondence may be summarized schematically as Eq.~(\ref{Sch}) and Eq.~(\ref{Cotton}).

The Schwarzian derivative and the Cotton tensor therefore appear to play
parallel roles as boundary invariants associated with the residual structure
of broken conformal gauge symmetry.

This viewpoint suggests that the AdS/CFT correspondence may admit a unified
interpretation in terms of boundary remnants of bulk conformal gauge geometry.

In particular, the Cotton tensor may play a role analogous to the Schwarzian derivative in the AdS$_2$/CFT$_1$ correspondence.
To clarify the boundary conformal structure more explicitly, let us consider
the AdS$_4$ metric in the Fefferman--Graham form \cite{FG, dHSS, Skenderis}
\begin{equation}
 ds^2=
 \frac{l^2}{z^2}
 \left[
 dz^2+g_{ij}(z,x)dx^idx^j
 \right].
\end{equation}

The metric is expanded near the boundary $z=0$ as
\begin{equation}
 g_{ij}(z,x)
 =
 g_{(0)ij}
 +z^2g_{(2)ij}
 +z^3g_{(3)ij}
 +\cdots.
\end{equation}

The induced metric on the boundary surface $z=\epsilon$ is
\begin{equation}
 \gamma_{ij}
 =
 \frac{l^2}{z^2}g_{ij}(z,x).
\end{equation}

The extrinsic curvature tensor is defined by
\begin{equation}
 K_{ij}
 =
 \frac12{\cal L}_n\gamma_{ij},
\end{equation}
where $n^\mu$ is the unit normal vector.

Using the Einstein-AdS$_4$ equation
\begin{equation}
 R_{\mu\nu}
 =
 -\frac{3}{l^2}g_{\mu\nu},
\end{equation}
we obtain
\begin{equation}
 g_{(2)ij}
 =
 -\left(
 R_{ij}[g_{(0)}]
 -\frac14R[g_{(0)}]g_{(0)ij}
 \right).
\end{equation}

The quantity in the bracket is the three-dimensional Schouten tensor
\begin{equation}
 P_{ij}
 =
 R_{ij}
 -\frac14Rg_{ij}.
\end{equation}

The mixed extrinsic curvature tensor becomes
\begin{equation}
 K^i{}_j
 =
 \frac1l\delta^i{}_j
 +\frac{z^2}{l}P^i{}_j
 -\frac{3z^3}{2l}g_{(3)}{}^i{}_j
 +\cdots.
\end{equation}

Thus, the first nontrivial correction to the AdS$_4$ extrinsic curvature
is governed by the boundary conformal geometry.

In three dimensions, the Weyl tensor vanishes identically,
and the conformal curvature is characterized by the Cotton tensor
\begin{equation}
 C_{ijk}
 =
 \nabla_kP_{ij}
 -\nabla_jP_{ik}.
\end{equation}

Therefore,
\begin{equation}
 \nabla_kK_{ij}
 -\nabla_jK_{ik}
 =
 \frac{z^2}{l}C_{ijk}
 +\cdots.
\end{equation}
The three-index Cotton tensor \(C_{ijk}\) is equivalent,
via dualization, to the symmetric Cotton tensor \(C_{ij}\)
appearing in Eq.~(27).
This structure may be regarded as a higher-dimensional analogue of the
Schwarzian derivative appearing in the AdS$_2$/CFT$_1$ correspondence.
This relation provides a direct geometrical link between
the asymptotic AdS$_4$ extrinsic geometry and the boundary
Cotton structure.
The present result suggests the correspondence
\[
K-\frac1l
\longleftrightarrow
\text{Schwarzian}
\qquad
(\text{AdS}_2/\text{CFT}_1),
\]
while in the AdS$_4$/CFT$_3$ case,
\[
\nabla K
\longleftrightarrow
\text{Cotton tensor}.
\]

This observation supports the interpretation that the boundary conformal
structure arises as a residual manifestation of the original conformal
gauge symmetry.

The present analysis suggests that the AdS$_4$/CFT$_3$ correspondence may be understood as a residual manifestation of the original conformal gauge symmetry.

\section{Remarks on AdS$_5$/CFT$_4$}

The situation becomes qualitatively different in the five-dimensional case.
In five dimensions, the straightforward extension of the gauge-theoretic
construction does not lead directly to the Einstein-Hilbert action linear
in the scalar curvature.
\begin{eqnarray}
\mathcal{L}_{gravity}&=&\epsilon^{ABCDEF}\epsilon^{\mu\nu\rho\sigma\lambda}R_{\mu\nu AB}R_{\rho\sigma CD}D_\lambda Z_EZ_F/l^2\nonumber\\
&=& \epsilon^{abcde}\epsilon^{\mu\nu\rho\sigma\lambda}R_{\mu\nu ab}R_{\rho\sigma cd}D_\lambda Z_e/l.
\label{RR}
\end{eqnarray}
Unlike the four-dimensional case,
the resulting action still contains bulk
higher-curvature contributions and does not reduce
to the Einstein--Hilbert form.


This suggests that the gauge-theoretic structure underlying the
AdS$_5$/CFT$_4$ correspondence may be qualitatively different from the
AdS$_4$/CFT$_3$ case.
Alternatively we had considered the linear $R$ term by
\begin{equation}
\mathcal{L}=\epsilon^{\mu\nu\rho\sigma\lambda}\epsilon^{abcde}R_{\mu\nu ab}D_\rho Z_cD_\sigma Z_d Z_e/l^4.
\label{R}
\end{equation}
 This structure appears less natural than the
four-dimensional case.

This suggests that the gravitational structure underlying AdS$_5$/CFT$_4$ may be fundamentally different from that of AdS$_4$/CFT$_3$.

In particular, the AdS$_4$/CFT$_3$ correspondence appears to be more naturally compatible with the gauge-theoretic formulation of gravity.
It is therefore conceivable that the AdS$_5$/CFT$_4$ correspondence should be regarded as an effective or emergent realization rather than a fundamental gauge-gravity structure.
The natural five-dimensional gauge-invariant density (\ref{RR}) should
not necessarily be judged by whether it reproduces the
five-dimensional Einstein--Hilbert action.  Equation (\ref{R}) can be
chosen if one wants the standard AdS$_5$ Einstein dynamics.  Rather,
the significance of (\ref{RR}) may be different: it appears to belong
to the descent sequence of a higher-dimensional Chern--Weil
density.  From this viewpoint, (\ref{RR}) may encode the conformal
anomaly structure of the four-dimensional boundary theory rather
than the ordinary bulk Einstein dynamics.

This suggests that in AdS$_5$/CFT$_4$ the conformal gauge-theoretic
origin of gravity may manifest itself not as the bulk
Einstein--Hilbert action, but rather through structures associated
with conformal anomalies and their descent relations.
In this respect, the situation may be contrasted with the
AdS$_4$/CFT$_3$ correspondence, where the Einstein--Hilbert action
emerges naturally from the symmetry breaking
SO$(4,2)\rightarrow$SO$(3,2)$.

From the present viewpoint, this may indicate that higher-dimensional
ingredients are fundamentally required once one goes beyond the
four-dimensional conformal gauge framework.
Thus (\ref{R}) and (\ref{RR}) may play complementary roles:
(\ref{R}) describes the effective Einstein--AdS$_5$ bulk dynamics,
whereas (\ref{RR}) points to the deeper anomaly/descent structure
that may underlie the holographic CFT$_4$ boundary.
The failure of the straightforward SO(4,2)-based
construction does not contradict the AdS$_5$/CFT$_4$
correspondence itself.
Rather, it may indicate that the latter relies on a
larger symmetry framework than SO(4,2) alone. It is also noteworthy that the standard AdS$_5$/CFT$_4$
correspondence is realized not by pure five-dimensional gravity,
but by type-IIB string theory on AdS$_5\times S^5$ \cite{Maldacena, Witten}.

\section{Conclusion}

We have discussed the AdS/CFT correspondence from the viewpoint of the
gauge-theoretic formulation of gravity.

In the AdS$_2$/CFT$_1$ case, the Schwarzian derivative was shown to emerge
naturally from the boundary extrinsic curvature of AdS$_2$ geometry.
The relation between the bulk Liouville geometry and the boundary projective
structure was clarified.
We further argued that the Brown--Henneaux type boundary central extension
should be interpreted not as a contradiction to the original gauge algebra,
but as an emergent asymptotic structure associated with broken conformal gauge
symmetry.
We then investigated the possible structure of the
AdS$_4$/CFT$_3$ correspondence, which is more directly connected
with the original four-dimensional formulation of gravity as a
broken phase of conformal gauge symmetry.

The essential observation is that, in the four-dimensional gauge-theoretic
formulation of gravity, the Einstein-Hilbert action with cosmological constant
emerges together with the total derivative structure
\[
\partial_\mu K^\mu .
\]
We argued that this boundary-sensitive structure naturally induces the
three-dimensional gravitational Chern--Simons term.
Its variation leads to the Cotton tensor, which plays the role of the
fundamental conformal invariant on the three-dimensional boundary.

This structure provides a natural higher-dimensional analogue of the
Schwarzian derivative appearing in AdS$_2$/CFT$_1$ correspondence:
\[
K-\frac1l
\longleftrightarrow
\{f,u\},
\]
for AdS$_2$/CFT$_1$, while
\[
\nabla_kK_{ij}-\nabla_jK_{ik}
\longleftrightarrow
C_{ijk},
\]
for AdS$_4$/CFT$_3$.

The Schwarzian derivative and the Cotton tensor therefore appear to play
parallel roles as boundary invariants associated with residual structures
of broken conformal gauge symmetry.
From this viewpoint, the AdS/CFT correspondence may admit a unified geometrical
interpretation in terms of boundary remnants of bulk conformal gauge geometry.

We also discussed the qualitative difference between AdS$_4$/CFT$_3$
and AdS$_5$/CFT$_4$.
In the straightforward five-dimensional extension of the gauge-theoretic
construction, higher-curvature structures arise naturally and the ordinary
Einstein-Hilbert action does not emerge in the same way as in four dimensions.
This suggests that the gauge-theoretic origin of AdS$_5$/CFT$_4$
may be qualitatively different from that of AdS$_4$/CFT$_3$.
The latter is naturally connected with gravity arising from the
 four-dimensional conformal gauge symmetry SO$(4,2)$ and its
breaking to SO$(3,2)$.
By contrast, the standard realization of AdS$_5$/CFT$_4$
is based on the higher-dimensional background
AdS$_5\times S^5$ of type-IIB string theory.
From this viewpoint, the success of AdS$_5$/CFT$_4$
may indicate that genuinely higher-dimensional,
string-inspired degrees of freedom become essential beyond the
four-dimensional conformal gauge framework considered here.
The present work is not intended to construct a detailed operator
dictionary of the AdS/CFT correspondence.
Rather, the main purpose is to clarify the geometrical origin of the
boundary conformal structures themselves from the viewpoint of broken
conformal gauge symmetry.

From this perspective, the AdS$_4$/CFT$_3$ correspondence may be
particularly natural within the gauge-theoretic formulation of gravity.
Further investigation of the boundary conformal dynamics and its relation to
bulk gauge geometry would therefore be an important direction for future investigation.
\appendix

\section{de Sitter Invariant Gravity}
Here we summarize the de Sitter invariant formulation of gravity for the reader's convenience.
Details may be found in the original papers \cite{FK1,FK2}.
\subsection{Four Dimensions}
(Anti-) de Sitter invariant gravity is based on the fact that 
\begin{itemize}
\item The maximal conformal symmetry of four-dimensional
electromagnetism and Yang--Mills theories is \(O(4,2)\).
\be
Z_1^2+......+Z_5^2+Z_6^2=0.
\ee
\item Gravity appears as it is broken to (anti-)de Sitter group
\be
Z_1^2+.....+Z_5^2=l^2
\ee
as the gauge field $\omega_{\mu AB}$ of the (anti) de Sitter gauge group, 

\begin{equation}
D_\mu Z_A=(\partial_\mu\delta_{AB}-\omega_{\mu AB})Z_B=
\left\{
\begin{array}{ll}
-\omega_{\mu a5}\,l
=
e_{\mu a},
& \qquad \mbox{if} ~~A=a,
\\[2mm]
0,
& \qquad \mbox{if}~~A=5 .
\end{array}
\right.
\end{equation}
Thus, the spin connection $\omega_{\mu ab}$ and the tetrad
$e_{\mu a}$ are treated on an equal footing as components of
the unified gauge field $\omega_{\mu AB}$.
The field strength is derived from the commutation relation
\be
i[D_\mu, D_\nu]=-R_{\mu\nu AB}S_{AB}/2
\ee
and
\be
R_{\mu\nu AB}=\partial_\mu\omega_{\nu AB}-\partial_\nu\omega_{\mu
AB}-\omega_{\mu AC}\omega_{\nu CB}+\omega_{\nu AC}\omega_{\mu CB}.
\ee
The gauge-invariant gravitational action is given by
\begin{eqnarray}
\mathcal{L}_{\rm gravity}
&=&
\frac{1}{16g^2}
\epsilon^{ABCDE}
\epsilon^{\mu\nu\rho\sigma}
\left(
\frac{Z_A}{l}
\right)
R_{\mu\nu BC}
R_{\rho\sigma DE}
\nonumber\\
&=&
\frac{1}{16g^2}
\epsilon^{abcd}
\epsilon^{\mu\nu\rho\sigma}
R_{\mu\nu ab}
R_{\rho\sigma cd}
\nonumber\\
&=&
\partial_\mu K^\mu
-
\frac{e}{16\pi G}
\left(
\mathring{R}
+
\frac{6}{l^2}
\right).
\label{K}
\end{eqnarray}

Here $A,B,\cdots$ ($a,b,\cdots$) run over $1-5$
($1-4$), respectively.

The curvature is decomposed as
\begin{equation}
R_{\mu\nu ab}
=
\mathring{R}_{\mu\nu ab}
+
\frac{1}{l^2}
e_{[\mu a}e_{\nu] b},
\end{equation}
where the summation on the right-hand side is restricted to the indices
$(1\text{--}4)$ and we have used
$-\omega_{\mu a5}l=e_{\mu a}.$
The important point is that the Einstein-Hilbert action with cosmological
constant emerges together with the total derivative term
$\partial_\mu K^\mu$.

\item de Sitter and anti-de Sitter case:

The difference between de Sitter and anti de Sitter case other than the signature of cosmological constant appears when gravity couples with matter \cite{FK1}.
\bea
\mathcal{L}_{spinor}&=&\epsilon^{ABCDE}\epsilon^{\mu\nu\rho\sigma}\overline{\psi}(S_{AB}\overset{\leftrightarrow}{D}_\mu/3!-\lambda(Z_A/l)D_\mu Z_B/4!)\psi \nonumber\\
&\times& D_\nu Z_CD_\rho Z_DD_\sigma Z_E ,
\eea
which leads to
\be
(e^{\mu a} \gamma^a\mathring{D}_\mu +m)\psi +le^{\mu a}\gamma^ae^{\nu b}R_{\mu\nu b5}\psi/2,
\ee
where $\mathring{D}_\mu$ is the covariant derivative acting only on the Lorentz indicies,
\be
\mathring{D}_\mu \psi=(\partial_\mu -i\omega_{\mu ab}S_{ab})\psi .
\ee
The mass of $\psi$ is the sum of two terms,
\be
m=\lambda-2i/l.
\ee
Thus the de Sitter case (real $l$) has the imaginary mass. 
\end{itemize}
\subsection{Two Dimensions}
The action is \cite{FK2}
\begin{equation}
I=\frac{1}{2}\int \epsilon^{ABC}R_{AB}\phi_C
\label{JT1}
\end{equation}
with $A,B,C=0,1,2$, which leads to JT gravity,
\begin{equation}
I=\int d^2x \sqrt{-g}(R-\Lambda)N
\label{JT2}
\end{equation}
with $\phi_2=N$.
In two dimensions, the symmetry breaking
\[
SO(2,2)\rightarrow SO(1,2)
\]
leads to AdS$_2$ gravity.
The resulting geometry is described by the Liouville structure
\[
e_{\mu a}=e^\chi\delta_{\mu a},
\]
and the boundary Schwarzian structure emerges from the extrinsic curvature as described in Sec.2.

In the original Fukuyama--Kamimura formulation,
the bulk AdS$_2$ gauge generators satisfy a closed algebra
without boundary contribution,
\begin{equation}
\{G[\xi],G[\eta]\}
=
G[[\xi,\eta]] .
\end{equation}

However, in the presence of an asymptotic AdS boundary,
the generator must be improved by adding a surface term,
\begin{equation}
\widetilde G[\xi]
=
G[\xi]+Q[\xi],
\end{equation}
so that its variation becomes well-defined under the imposed
boundary conditions.

The Poisson bracket of the improved generators then takes the form
\begin{equation}
\{\widetilde G[\xi],\widetilde G[\eta]\}
=
\widetilde G[[\xi,\eta]]
+
K[\xi,\eta],
\end{equation}
where \(K[\xi,\eta]\) is a boundary contribution.

For asymptotic conformal transformations,
the improved generators become the Virasoro generators \(L_n\),
and the algebra takes the form
\begin{equation}
i\{L_m,L_n\}
=
(m-n)L_{m+n}
+
\frac{c}{12}m(m^2-1)\delta_{m+n,0}.
\label{algebra}
\end{equation}

Thus the central extension originates not from the original bulk
gauge algebra itself, but from the boundary contribution associated
with asymptotic AdS boundary conditions.
In the Brown--Henneaux normalization \cite{BH}, the central charge is
\begin{equation}
c=\frac{3l}{2G}.
\label{center}
\end{equation}

Thus the central charge does not belong to the original bulk gauge algebra.
It appears only after imposing asymptotic AdS boundary conditions and adding
the corresponding surface charges.

Thus, in both two and four dimensions, the gauge-theoretic formulation
naturally leads to characteristic boundary structures associated with
conformal geometry.
The dimensional dependence of this mechanism becomes important when one considers higher-dimensional extensions.
This suggests that the gauge-theoretic structure underlying the
AdS$_5$/CFT$_4$ correspondence may be qualitatively different from the
AdS$_4$/CFT$_3$ case.

\end{document}